\documentclass[aps,pre,twocolumn,superscriptaddress,showpacs,showkeys]{revtex4}

\usepackage{natbib}
\usepackage{dcolumn}
\usepackage{amsmath}
\usepackage{amssymb}
\usepackage{graphicx}
\usepackage{rotating}
\usepackage{multirow}

\begin{document}

\title{Benchmarks for testing community detection algorithms on directed and weighted graphs with overlapping communities}

\author{Andrea Lancichinetti}
\affiliation{Complex Networks Lagrange Laboratory (CNLL),
Institute for Scientific Interchange (ISI), Viale S. Severo 65, 10133, Torino, Italy}

\author{Santo Fortunato}
\affiliation{Complex Networks Lagrange Laboratory (CNLL),
Institute for Scientific Interchange (ISI), Viale S. Severo 65, 10133, Torino, Italy}

\begin{abstract}

Many complex networks display a mesoscopic structure with groups of nodes sharing many links
with the other nodes in their group and comparatively few with nodes of different groups. This feature is known 
as community structure and encodes precious information about the 
organization and the function of the nodes. Many algorithms have been proposed but it is not yet clear how 
they should be tested. Recently we have proposed a general class of undirected and unweighted 
benchmark graphs, with heterogenous distributions
of node degree and community size. An increasing 
attention has been recently devoted to develop algorithms able to consider the direction and the 
weight of the links, which require suitable benchmark graphs for testing.
In this paper we extend the basic 
ideas behind our previous benchmark to generate directed and weighted networks with built-in community structure.
We also consider the possibility that nodes belong to more communities, a feature
occurring in real systems, like social networks. As a practical application, we show how
modularity optimization performs on our new benchmark. 
 
\end{abstract}

\pacs{89.75.-k, 89.75.Hc}
\keywords{Networks, community structure}
\maketitle

\section{Introduction}

Complex systems are characterized by a division in subsystems, which in turn contain other subsystems in a hierarchical fashion.
Herbert A. Simon, already in 1962, pointed out that such hierarchical organization plays a crucial role both in the generation and in the
evolution of complex systems~\cite{simon62}. Many complex systems can be described as graphs, or networks,
where the elementary parts of a system and their mutual interactions are nodes and links, respectively~\cite{Newman:2003,vitorep}. In a network,
the subsystems appear as subgraphs with a high density of internal links, which are loosely connected to each other. 
These subgraphs are called communities and occur in a wide variety of networked systems~\cite{Girvan:2002,miareview}.
Communities reveal how a network is internally organized, and indicate the presence of special
relationships between the nodes, that may not be easily accessible from direct empirical tests.
Communities may be groups of related individuals 
in social networks~\cite{Girvan:2002, Lusseau:2005}, sets of Web pages dealing with the same topic~\cite{Flake:2002}, 
biochemical pathways in metabolic networks~\cite{Guimera:2005,palla}, etc. 

For these reasons, detecting communities in networks has become a fundamental problem in network science. Many methods have been developed,
using tools and techniques from disciplines like physics, biology, applied mathematics, 
computer and social sciences. However, there is no agreement yet 
about a set of reliable algorithms, that one can use in applications. The main reason is that current techniques have 
not been thoroughly tested. Usually, when a new method is presented, it is applied to a few simple benchmark graphs,
artificial or from the real world, which have a known community structure. The most used benchmark is a class of graphs introduced
by Girvan and Newman~\cite{Girvan:2002}. Each graph consists of $128$ nodes, which are divided into four groups of $32$: 
the probabilities of the existence of a link between a pair of nodes of the same group and of 
different groups are $p_{in}$ and $p_{out}$, respectively. 
This benchmark is a special case of the {\it planted $\ell$-partition model}~\cite{condon01}. However,  
it has two drawbacks: 1) all nodes have the same
expected degree; 2) all communities have equal size. These features are unrealistic, as complex networks are known to be characterized by
heterogeneous distributions of degree~\cite{albert99,Newman:2003,vitorep} and community sizes~\cite{palla,guimera03,arenasrev,clausetfast,lancichinetti09}.
In a recent paper~\cite{lancichinetti08}, we have introduced a new class of benchmark graphs, 
that generalize the benchmark by Girvan and Newman by introducing power law distributions of degree and community size. 
Most community detection algorithms perform very well on the benchmark 
by Girvan and Newman, due to the simplicity of its structure.
The new benchmark, instead, poses a much harder test to algorithms, and makes it easier to disclose their limits.

Most research on community detection focuses on the simplest case of undirected and unweighted graphs, as the problem 
is already very hard. However, links of networks from the real world are often directed and carry weights,
and both features are essential to understand their function~\cite{leicht08,barrat04}. Moreover, in real graphs communities are 
sometimes overlapping~\cite{palla}, i. e. they share vertices. This aspect, frequent in certain types of systems, like social networks,
has received some attention in the last years~\cite{baumes05,zhang07,nepusz08,lancichinetti09}. 
Finding communities in networks with directed and weighted edges and possibly overlapping communities is highly non-trivial. 
Many techniques working on undirected graphs, for instance, cannot be extended to include link direction. 
This implies the need of new approaches to the problem. In any case, once a method is designed, it is important 
to test it against reliable benchmarks. Since the new benchmark of Ref.~\cite{lancichinetti08} is defined for undirected and unweighted graphs,
we extend it here to the directed and weighted cases. For any type of benchmark, we will include the possibility to have 
overlapping communities. Sawardecker et al. have recently proposed a different benchmark with overlapping communities
where the probability that two nodes are linked grows with the number of communities both nodes belong to~\cite{sawardecker09}.
 
Our algorithms to create the benchmark graphs have a computational complexity which grows
linearly with the number of links and reduce considerably the fluctuations of specific realizations of the graphs,
so that they come as close as possible 
to the type of structure described by the input parameters. We use our benchmark to make some 
testing of modularity optimization~\cite{newman04}, which is well defined in the case of directed and weighted networks~\cite{arenas07c}.

In Section~\ref{sec1} we describe the algorithms to create the new benchmarks. Tests are presented in Section~\ref{sec2}.
Conclusions are summarized in Section~\ref{sec3}.

\section{The benchmark}
\label{sec1}

We start by presenting the algorithm to build the benchmark for undirected graphs with overlaps between communities.
Then we extend it to the case of weighted and directed graphs.

\subsection{Unweighted benchmark with overlapping nodes}
\label{sec21}

The aim of this section is to describe the algorithm to generate undirected and unweighted benchmark graphs, 
where each node is allowed to have memberships in more communities. The algorithm consists of the following steps:

\begin{enumerate}
\item We first assign the number $\nu_i$ of memberships of node $i$, i.e. the number of 
communities the node belongs to. Of course, if each node has only one membership, we recover the  
benchmark of Ref.~\cite{lancichinetti08}; in general we can assign the number of memberships according to a certain distribution. 
Next, we assign the degrees $\{k_i\}$ by drawing $N$ random numbers from a power law distribution~\cite{power} with 
exponent $\tau_1$. We also introduce the $\textit{topological mixing parameter}$ $\mu_t$: $k_i^{(in)} = (1- \mu_t ) k_i  $
is the internal degree of the node $i$, i. e. the number of 
neighbors of node $i$ which have at least one membership in common with $i$. 
In this way, the internal degree is a fixed fraction of the total degree 
for all the nodes. Of course, it is straightforward to generalize the algorithm to implement a different 
rule (one can introduce a non linear functional dependence, individual mixing parameters, etc.). 
     
 \item The community sizes $\{s_\xi\}$ are assigned by drawing random numbers from 
another power law with exponent $\tau_2$. Naturally, the sum of the community sizes must equal the 
sum of the node memberships, i. e. $ \sum_\xi s_\xi = \sum_i \nu_i $. Furthermore 
$ s_{max} = \max\{s_\xi\} \leqslant N$ and $\nu_{max} = \max\{\nu_i\} \leqslant n_c$, 
where $N$ is the number of nodes and $n_c$ the number of communities. At this point, we have to 
decide which communities each node should be included into. This is equivalent to generating a bipartite 
network where the two classes are the $n_c$ communities and the $N$ nodes; each community $\xi$ has $s_\xi$ links, whereas  
each node has as many links as its memberships $\nu_i$ (Fig.~\ref{bipart}). 
\begin{figure} [ht]
\centering
\includegraphics[width=0.7\columnwidth]{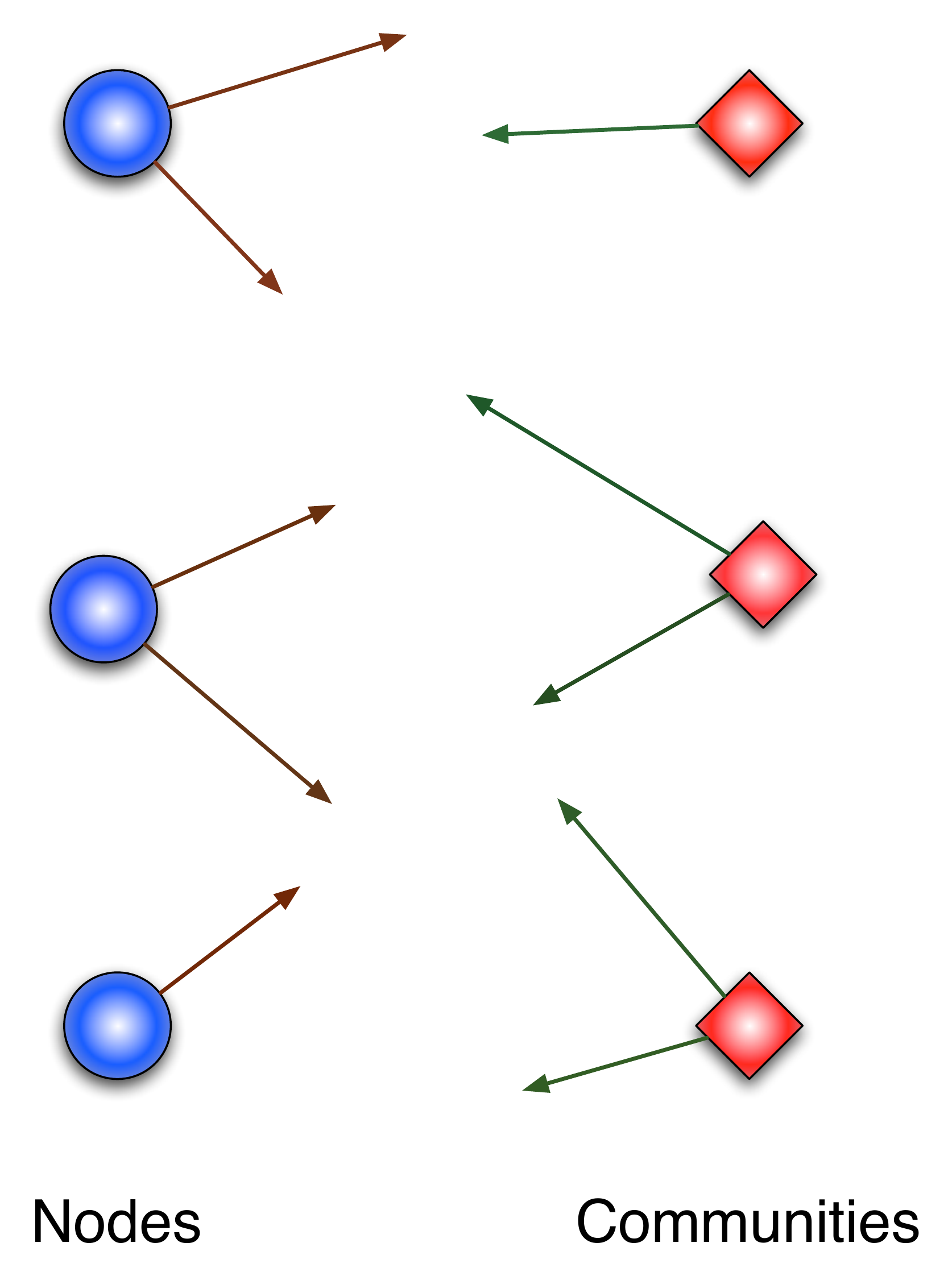} 
\caption {\label{bipart} Schematic diagram of the bipartite graph used to assign nodes to their communities. 
Each node has as many stubs as the number of communities it belongs to, whereas the number of stubs of each community matches the size
of the community. The memberships are assigned by joining the stubs on the left with those on the right.} 
\end{figure}
The network can be easily generated with the configuration model~\cite{molloy95}. 
To build the graph, it is important to take into account the constraint 
\begin{equation}
\label{constraint_weak}
\sum_{i\rightarrow \xi} s_\xi\geqslant k_i^{(in)}, \hskip1cm \forall i, 
\end{equation}
where the sum is relative to the communities including node $i$. This condition means that each node 
cannot have an internal degree larger than the highest possible number of nodes it can be connected to 
within the communities it stays in. We perform a 
rewiring process for the bipartite network until the constraint is satisfied. For some choices of the input parameters,
it could happen that, after some 
iterations, the constraint is still unsatisfied. In this case one can change the sizes of the communities, by merging some of them, 
for instance. It turns out that this is not necessary in most situations and that, when it is, the perturbations introduced in the
community size distributions are not too large.
In general, it is convenient to start with a distribution of community sizes such that 
$s_{min} \geqslant  k_{min}^{(in)}$ and $s_{max} \geqslant k_{max}^{(in)}$. 
 
So far we assigned an internal degree to each node but it has not been specified how many links should 
be distributed among the communities of the node. Again, one can follow several recipes; we chose the simple equipartition 
$k_i(\xi)=  k_i^{(in)} / \nu_i$, where $k_i(\xi)$ is the number of links which $i$ shares in community $\xi$,  
provided that $i$ holds membership in $\xi$. Some adjustments may be necessary to assure 
\begin{equation}
\label{constraint_strong}
(s_\xi)_{i\rightarrow\xi} \geqslant k_i(\xi) \,\,\, \forall i,
\end{equation}
which is the strong version of Eq.~\ref{constraint_weak}.

 \item Before generating the whole network, we start generating $n_c$ subgraphs, one for each community. 
In fact, our definition of community $\xi$ is nothing but a random subgraph of $s_{\xi}$ nodes with degree 
sequence $\{k_i(\xi)\}$, which can be built via the configuration model, with a rewiring procedure to 
avoid multiple links. Note that Eq.~\ref{constraint_strong} is necessary to generate the configuration 
model, but in general not sufficient. For one thing, we need $\sum_i k_i(\xi) $ to be even. 
This might cause a change in the degree 
sequence, which is generally not appreciable. Once each subgraph is built, we obtain a graph divided in components. 
Note that because of the overlapping nodes, some components may be connected to each other, 
and in principle the whole graph might be connected. Furthermore, 
if two nodes belong simultaneously to the same two (or more) communities, the procedure may set more than one link
between the nodes. A rewiring strategy similar to that described below suffices to avoid this problem.
  
\item The last step of the algorithm consists in adding the links external to the communities. To do this, let us consider the 
degree sequence $\{k_i^{(ext)}\}$, where simply $k_i^{(ext)} = k_i - k_i^{(in)} = \mu_t k_i$. We want to 
insert randomly these links in our already built network without changing the internal degree sequences. 
In order to do so, we build a new network $\mathcal{G}^{(ext)}$ of $N$ nodes with 
degree sequence $\{k_i^{(ext)}\}$, and we perform a rewiring process each time we encounter a link 
between two nodes which have at least one membership in common (Fig.~\ref{rewire}), since we are supposed to join only 
nodes of different communities at this stage. 
\begin{figure} [ht]
\centering
\includegraphics[width=0.7\columnwidth]{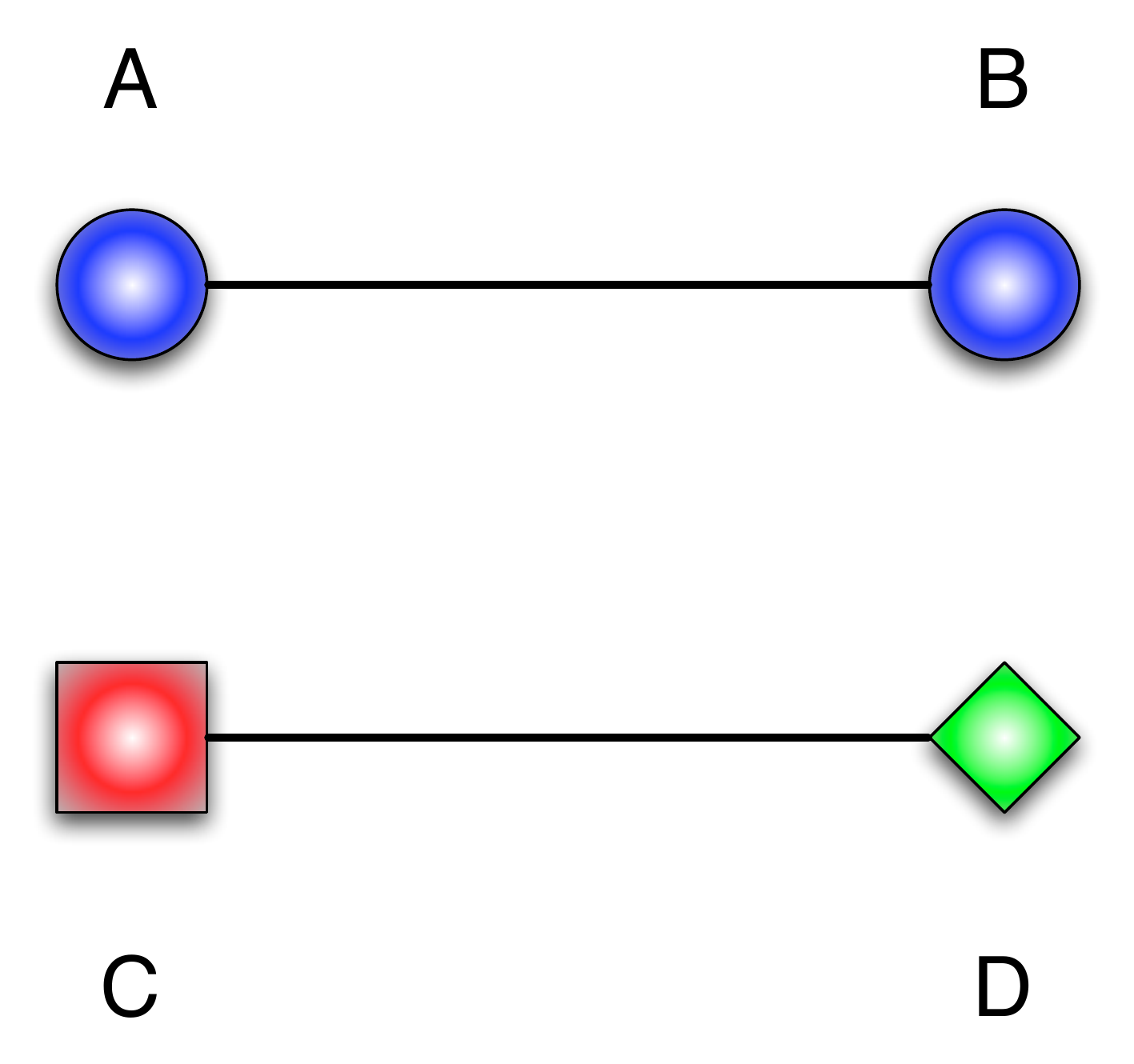} 
\includegraphics[width=0.7\columnwidth]{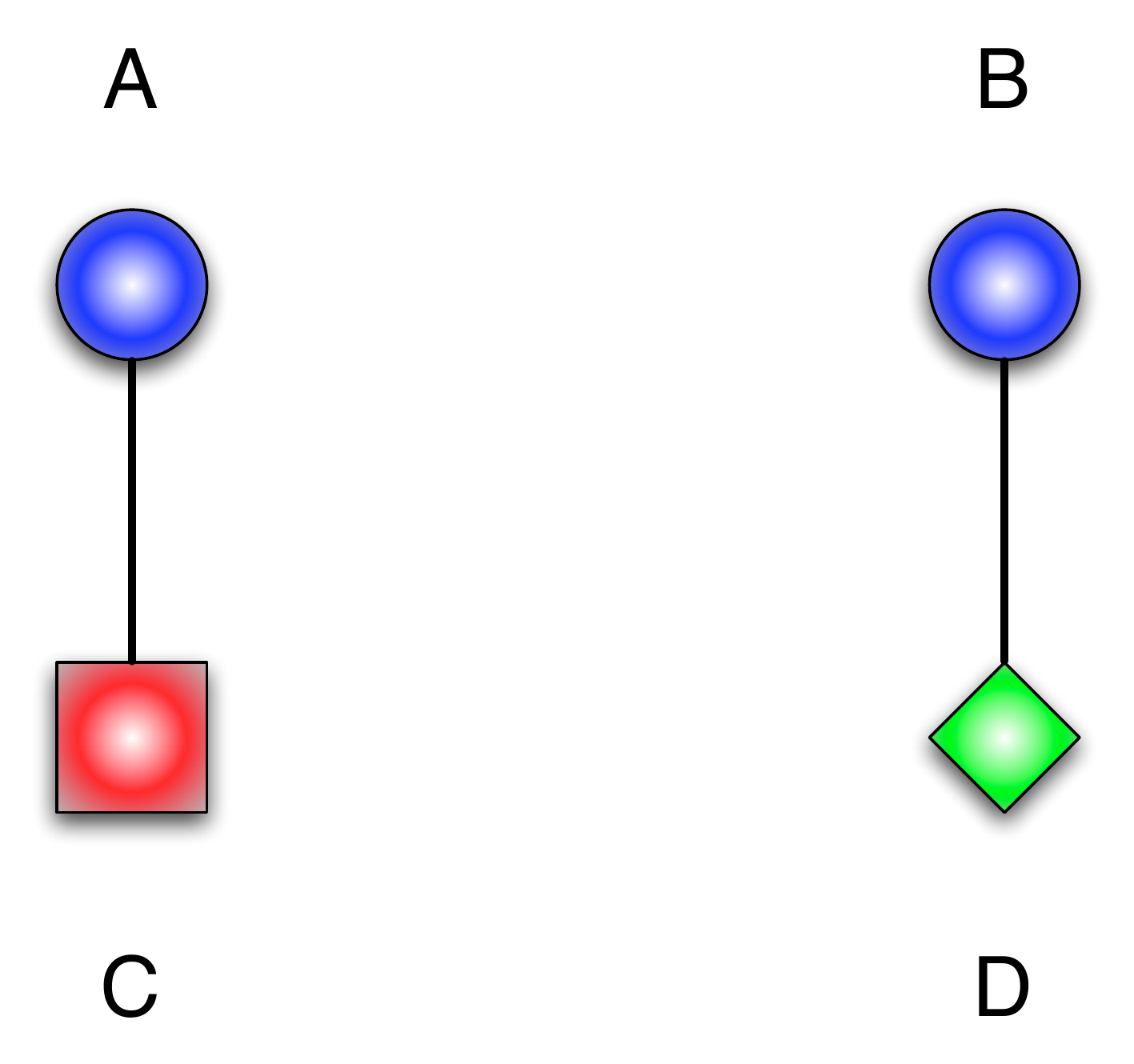} 
\caption {\label{rewire} Scheme of the rewiring procedure necessary to build the graph $\mathcal{G}^{(ext)}$, which 
includes only links between nodes of different communities. (Top) If two nodes ($A$ and $B$) 
with a common membership are neighbors, their link is 
rewired along with another link joining two other nodes $C$ and $D$, where
$C$ does not have memberships in common with $A$, and $D$ is a neighbor of $C$ not connected to $B$.
In the final configuration (bottom), the degrees of all nodes are preserved, and the number of links between 
nodes with common memberships has decreased by one (since $A$ and $B$ are no longer connected), or 
it has stayed the same (if $B$ and $D$, which are now neighbors, have common memberships).}
\end{figure}
Let us assume that $A$ and $B$ are in the same 
community and that they are linked in $\mathcal{G}^{(ext)}$; we pick a node $C$ which does not share any 
membership with $A$, and we look for a neighbor of $C$ (call it $D$) which is not neighbor of $B$. 
Next, we replace the links $A-B$ and $C-D$ with the new links 
$A-C$ and $B-D$. This rewiring procedure can decrease the number of internal links of $\mathcal{G}^{(ext)}$ 
or leaving it unchanged (this happens only when $B$ and $D$ have one membership in common) but it cannot 
increase it. This means that after a few sweeps over all the nodes we reach a steady state where the 
number of internal links is very close to zero (if no node has $k_i \sim N$, 
the internal links of $\mathcal{G}^{(ext)}$ are just a few and one sweep is sufficient). 
Fig.~\ref{Fig_internal_degree_g_ext} 
shows how the number of internal links decreases during the rewiring procedure.
Finally, we have to superimpose $\mathcal{G}^{(ext)}$ on the previous one.
\end{enumerate}

In our previous work about benchmarking~\cite{lancichinetti08}, we discussed 
the dispersion of the internal degree around the fixed value $k_i^{(in)}$. 
In this case, if the number of internal links of $\mathcal{G}^{(ext)}$ goes to zero, the only reason 
not to have a perfectly sharp function for the distribution of the mixing parameters of the nodes in 
specific realizations of the new benchmark is a round-off problem, i.e. the problem of rounding integer numbers. 
\begin{figure} [ht]
\centering 
\includegraphics[height=0.4 \textwidth, width=0.5 \textwidth]{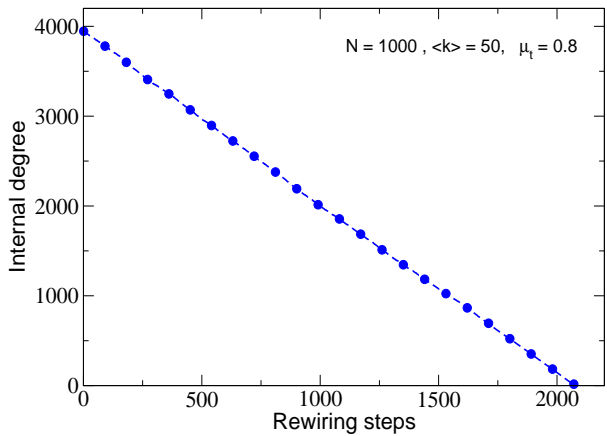} 
\caption{\label{Fig_internal_degree_g_ext}Number of internal links of $\mathcal{G}^{(ext)}$ as a function of the rewiring steps. 
The network has $1000$ nodes, and an average degree $\langle k \rangle =50$. Since the mixing parameter 
is $\mu_t=0.8$ and there are $10$ equal-sized communities, at the beginning each node has an expected 
internal degree in $\mathcal{G}^{(ext)}$ $k_i^{(in)}= 0.8 * 50 * 1/10 = 4$, so the total internal degree 
is around $4000$. After each rewiring step, the internal degree either decreases by $2$, or it does not change. 
In this case, less then $2100$ rewiring steps were needed.} 
\end{figure} 
 
Other benchmarks, like that by Girvan and Newman, are based on a similar definition of communities, expressed 
in terms of different probabilities for internal and external links. One may wonder what is 
the connection between our benchmark and the others. It is not difficult to compute an approximation of 
how the probability of having a link between two nodes in the same community depends on the mixing parameter $\mu_t$.
 
In the configuration model, the probability to have a connection between nodes $i$ and $j$ with $k_i$ 
and $k_j$ links respectively is approximately $p_{ij}= \frac{k_i k_j}{2m}$, provided that $k_i \ll 2m$ 
and $k_j \ll 2m$.  If the approximation holds, our prescription to assign $k_i(\xi)$ allows us to compute 
the probability that $i$ and $j$ get a link in the community $\xi$:
\begin{equation}
p_{ij}(\xi) \simeq \frac{k_i(\xi) k_j(\xi)}{2m_{\xi}}  = (1 - \mu_t)^2 \frac{1}{\nu_i \nu_j} \,\, \frac{k_i k_j}{2m_{\xi}}, 
\end{equation}
where $2m_{\xi}= \sum_i k_i(\xi)$ is the number of internal links in the community (we recall that $\nu_i$ is 
the number of memberships of node $i$). If $i$ and $j$ share a number $\nu_{ij}$ of memberships and all the 
respective $p_{ij}(\xi)$ are small, the probability that they get a link somewhere can be approximated with the 
sum over all the common communities. The final result is
\begin{equation}
p_{ij} \simeq(1 - \mu_t)^2 k_i k_j \frac{\nu_{ij}}{\nu_i \nu_j} \,\,  \langle  \frac{1}{2m_{\xi}} \rangle _{\xi}, 
\end{equation}
where $\langle  \frac{1}{2m_{\xi}} \rangle_{\xi}   = 1/ \nu_{ij} \sum_{\xi} 1/2m_{\xi}$, and $\xi$ runs 
only over the common memberships of the nodes.

On the other hand, if $i$ and $j$ do not share any membership, the probability to have a link between them is:
\begin{equation}
p_{ij} \simeq \frac{k_i^{(ext)} k_j^{(ext)}}{2m^{(ext)}}  =  \mu_t \,\, \frac{k_i k_j}{2m} 
\end{equation}
where $2m^{(ext)}= \sum_i k_i^{(ext)} = \mu_t \sum_i k_i$ is the number of external links in the network. 
The equation holds only if the rewiring process does not affect too much the probabilities, i.e. 
if the communities are small compared to the size of the network.

These results are based on some assumptions which are likely to be not exactly, but only approximately valid. Anyway, carrying out the right 
calculation is far from trivial and surely beyond the scope of this paper.

\begin{figure} [ht]
\centering 
\includegraphics[height=0.4 \textwidth, width=0.5 \textwidth]{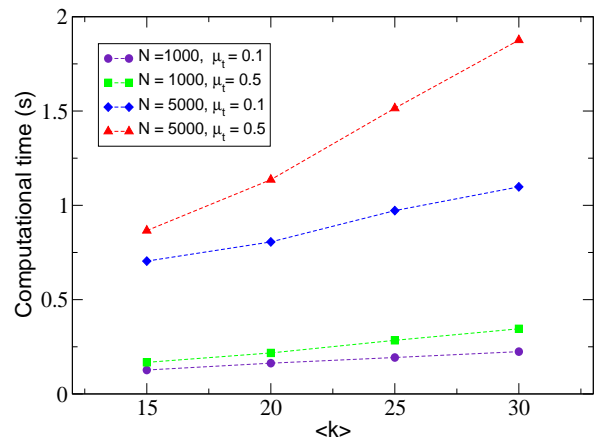} 
\caption{\label{comp}Computational time to build the unweighted benchmark as a function of the average degree. 
We show the results for networks of $1000$ and $5000$ nodes. $\mu_t$ was set equal to $0.1$ and $0.5$ 
(the latter requires more time for the rewiring process). Note that between the two upper lines and the 
lower ones there is a factor of about 5, as one would expect if complexity is linear in the number of links $m$.} 
\end{figure}

We conclude this section with a remark about the complexity of the algorithm. The configuration model 
takes a time growing linearly with the number of links $m$ of the network. If the rewiring procedure takes only a few 
iterations, like it happens in most instances, the complexity of the algorithm is $O(m)$ (Fig.~\ref{comp}).

\subsection{Weighted networks}

In order to build a weighted network, we first generate an unweighted network with a given topological mixing parameter 
$\mu_t$ and then we assign a positive real number to each link. 

To do this we need to specify two other parameters, $\beta$ and $\mu_w$. The parameter $\beta$ is used to assign 
a strength $s_i$ to each node, $s_i = k_i^{\beta}$; such power law relation between the strength and the degree
of a node is frequently observed in real weighted networks~\cite{barrat04}.
The parameter $\mu_w$ is used to assign the internal strength 
$s_i^{(in)}= (1-\mu_w) \,\, s_i$, which is defined as the sum of the weights of the links between node $i$ 
and all its neighbors having at least one membership in common with $i$. The problem is equivalent to 
finding an assignment of $m$ positive numbers $\{ w_{ij} \}$ such to minimize the following function:
\begin{equation}
\text{Var}(\{ w_{ij} \}) = \sum_i  (s_i - \rho_i)^2 + (s_i^{(in)} - \rho_i^{(in)})^2 + (s_i^{(ext)} - \rho_i^{(ext)})^2.
\label{variance}
\end{equation}
Here $s_i$ and $s_i^{(in,ext)}$ indicate the strengths which we would like to assign, i.e. 
$s_i = k_i^{\beta}$,  $s_i^{(in)}= (1-\mu_w) \,\, s_i$, $s_i^{(out)}=\mu_w \,\, s_i$; 
$\{\rho_i^*\}$ are the total, internal and external strengths of node $i$ defined through 
its link weights, i.e. $\rho_i = \sum_j w_{ij} $, 
$\rho_i^{(in)}= \sum_j w_{ij} \,\, \kappa(i, j) $, $\rho_i^{(out)}= \sum_j w_{ij} \,\, (1-\kappa(i, j) )$, 
where the function $\kappa(i, j) = 1$ if nodes $i$ and $j$ share at least one membership, and $\kappa(i, j) =0$ otherwise.

We have to arrange things so that $s_i$ and $s_i^{(in,ext)}$ are consistent 
with the $\{\rho_i^*\}$. For that we need a fast algorithm 
to minimize $\text{Var}(\{ w_{ij} \})$. We found that the greedy algorithm 
described below can do this job well enough for the cases of our interest.

\begin{figure} [!h]
\centering
\includegraphics[height=0.4 \textwidth, width=0.5 \textwidth]{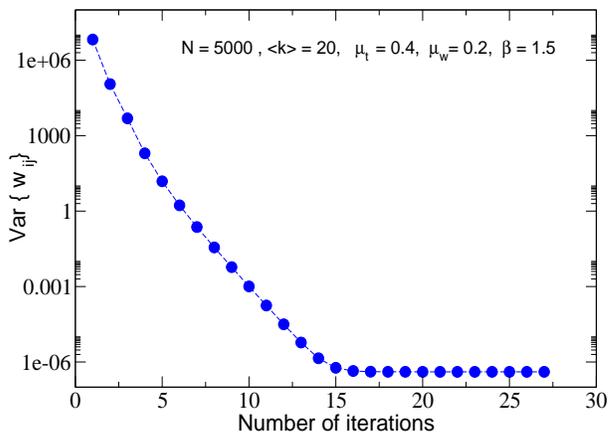} 
\caption{Value of $\text{Var}(\{ w_{ij} \})$ (Eq.~\ref{variance}) after each update. Each point corresponds to one sweep over all the nodes.}
\label{Figvar}
\end{figure}

\begin{enumerate}

\item At the beginning $w_{ij}=0$, $\forall i,j$, so all the $\{\rho_i^*\}$ are zero.

\item We take node $i$ and increase the weight of each of its links by an amount $u_i = \frac{s_i - \rho_i}{k_i}$, 
where $\rho_i$ indicates the sum of the links' weights resulting from the previous step, i. e.
before we increment them. In this way, since initially 
$\{\rho_i^*\}=0$, the weights of the links of $i$ after the first step take the (equal for all) value $\frac{s_i}{k_i}$,
and $\rho_i = s_i$ by construction, condition that is maintained along the whole procedure. 
We update $\{\rho_i^*\}$ for the node $i$ and its neighbors.

\item Still for node $i$ we increase all the weights $w_{ij}$ by an amount
$\frac{s_i^{(in)} - \rho_i^{(in)}}{k_i^{(in)}}$ if $\kappa(i, j) = 1$ and by an amount
$-\frac{s_i^{(in)} - \rho_i^{(in)}}{k_i^{(ext)}}$ if $\kappa(i, j) = 0$.
Again we update $\{\rho_i^*\}$ for the node $i$ and its neighbors.
These two steps assure to set the contribute of node $i$ in $\text{Var}(\{ w_{ij} \})$ to zero.

\item We repeat steps (2) and (3) for all the nodes. Two remarks are in order. First, we want each weight 
$w_{ij} >  0$; so we update the weights only if this condition is fulfilled. 
Second, the contribute of 
the neighbors of node $i$ in $\text{Var}(\{ w_{ij} \})$ will change and, of course, it can increase or 
decrease. For this reason, we need to iterate the procedure several times until a steady state is reached, 
or until we reach a certain value. With our procedure the value of $\text{Var}(\{ w_{ij} \})$ decreases at least 
exponentially with the number of iterations, consisting in sweeps over all network links.
(Fig. \ref{Figvar}).

\end{enumerate}

\begin{figure} [ht]
\centering 
\includegraphics[height=0.4 \textwidth, width=0.5 \textwidth]{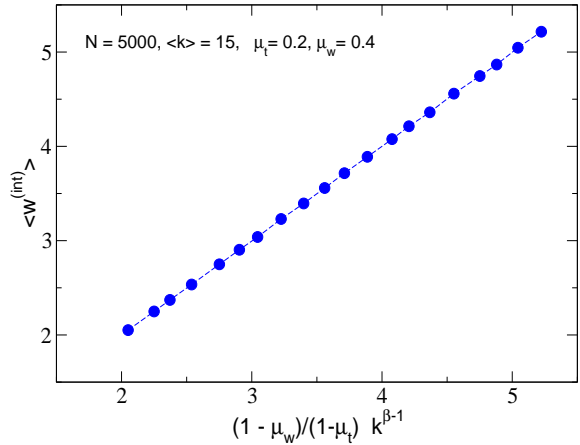} 
\caption{\label{avwei} The average weight of an internal link for a node depends on its degree according to 
Eq.~\ref{Eqaverage}. The correlation plot in the figure, relative to a network of $5000$ nodes, confirms the result.} 
\end{figure}

For the distribution of the weights $w_{ij}$, we expect the averages
$\langle w_{i}  ^{(int)} \rangle =  1/k_i^{(in)}	\sum_j w_{ij} \kappa(i, j) = s_i^{(in)} / k_i^{(in)}$ and 
$\langle w_{i}  ^{(ext)} \rangle = s_i^{(ext)} / k_i^{(ext)} $. Note that these expressions can be 
related to the mixing parameters in a simple way (Fig.~\ref{avwei}):

\begin{equation}
\label{Eqaverage}
\langle w_{i}  ^{(int)} \rangle = \frac{1 - \mu_w}{1- \mu_t} \,\,k_i^{\beta -1}  \qquad 
\text{and} \qquad \langle w_{i}  ^{(ext)} \rangle = \frac{\mu_w}{\mu_t} \,\,k_i^{\beta -1}.
\end{equation}

\begin{figure} [!h]
\centering 
\includegraphics[height=0.4 \textwidth, width=0.5 \textwidth]{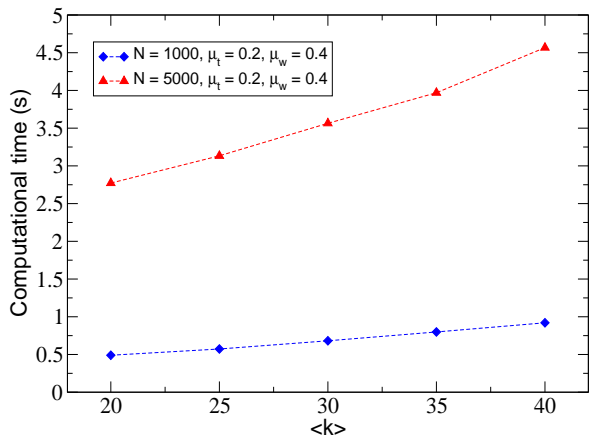} 
\caption{\label{comp2} Computational time to build the weighted benchmark as a function of the average degree. 
We show the results for networks of $1000$ and $5000$ nodes. $\mu_t$ was set equal to $0.2$ and $\mu_w$ to $0.4$.} 
\end{figure}

Since $\text{Var}(\{ w_{ij} \})$ decreases exponentially, 
the number of iterations needed to reach convergence has a slow dependence on the size of the network so it does not contribute 
much to the total complexity, which remains $O(m)$ (Fig.~\ref{comp2}).

\subsection{Directed networks}

It is quite straightforward to generalize the previous algorithms to generate directed networks. 
Now, we have an indegree sequence $\{y_i\}$ and an outdegree sequence $\{z_i\}$ but we 
can still go through all the steps of the construction of the benchmark for undirected networks with just some slight modifications. 
In the following, we list what to change in each point of the corresponding list in Section~\ref{sec21}.

\begin{enumerate}
  \item We decided to sample the indegree sequence from a power law and the outdegree sequence from a $\delta$-distribution 
(with the obvious constraint $\sum_i y_i =  \sum_i z_i$). We need to define the internal in- and outdegrees $y_i(\xi)$ and $z_i(\xi)$ 
with respect to every community $\xi$, which can be done by introducing two mixing parameters. For simplicity one can set them equal. 

  \item It is necessary that Eq.~\ref {constraint_strong} holds for both $\{y_i\}$ and $\{z_i\}$.
  \item We need to use the configuration model for directed networks, and the condition that $\sum_i k_i(\xi) $ should be 
even is replaced by $\sum_i y_i(\xi) =  \sum_i z_i(\xi) $; because of this condition 
it might be necessary to change $y_i(\xi)$ and/or $z_i(\xi)$. We decided to modify only $z_i(\xi)$, whenever necessary.
  
  \item The rewiring procedure can be done by preserving both distributions of indegree and outdegree, 
for instance, by adopting the following scheme: 
before rewiring, $A$ points to $B$ and $D$ to $C$; after rewiring, $A$ points to $C$ and $D$ to $B$.
  
\end{enumerate}

In order to generate directed and weighted networks, we use the following relation between the strength
$s_i$ of a node and its in- and outdegree: $s_i = (y_i + z_i)^{\beta}$. Given a node $i$,
one considers all its neighbors, regardless of the link directions 
(note that $i$ may have
the same neighbor counted twice if the link runs in both directions). 
Otherwise, the procedure to insert weights is equivalent.

In directed networks, the directedness of the links may reflect some interesting structural information that 
is not present in the corresponding undirected version of the graph. 
For instance there could be flows, represented by many links with the same direction 
running from one subgraph to another: such subgraphs 
might correspond to important classifications of the nodes. 
Our directed benchmark is based on the balance between the numbers of internal and external links,
and it does not seem suitable to generate graphs with flows. However, this is not true: flows can be generated 
by introducing proper constraints on the number of incoming and outgoing links of the communities.  

Suppose we want to generate a network with two communities only, where the nodes of 
community $1$ point to nodes of community $2$ but not vice versa and there are a few random connections among 
nodes in the same community. We could use our algorithm in this way: first we build separately the two 
subgraphs; then we set $y_i^{(ext)} \simeq 0$ for nodes in the community $1$ and 
$z_i^{(ext)} \simeq 0$ for nodes in community $2$ and build $\mathcal{G} ^{(ext)}$. If there are more 
communities, one first builds as many subgraphs as necessary and then links them according to the desired flow patterns.

\begin{figure} [ht]
\centering
\includegraphics[width=\columnwidth]{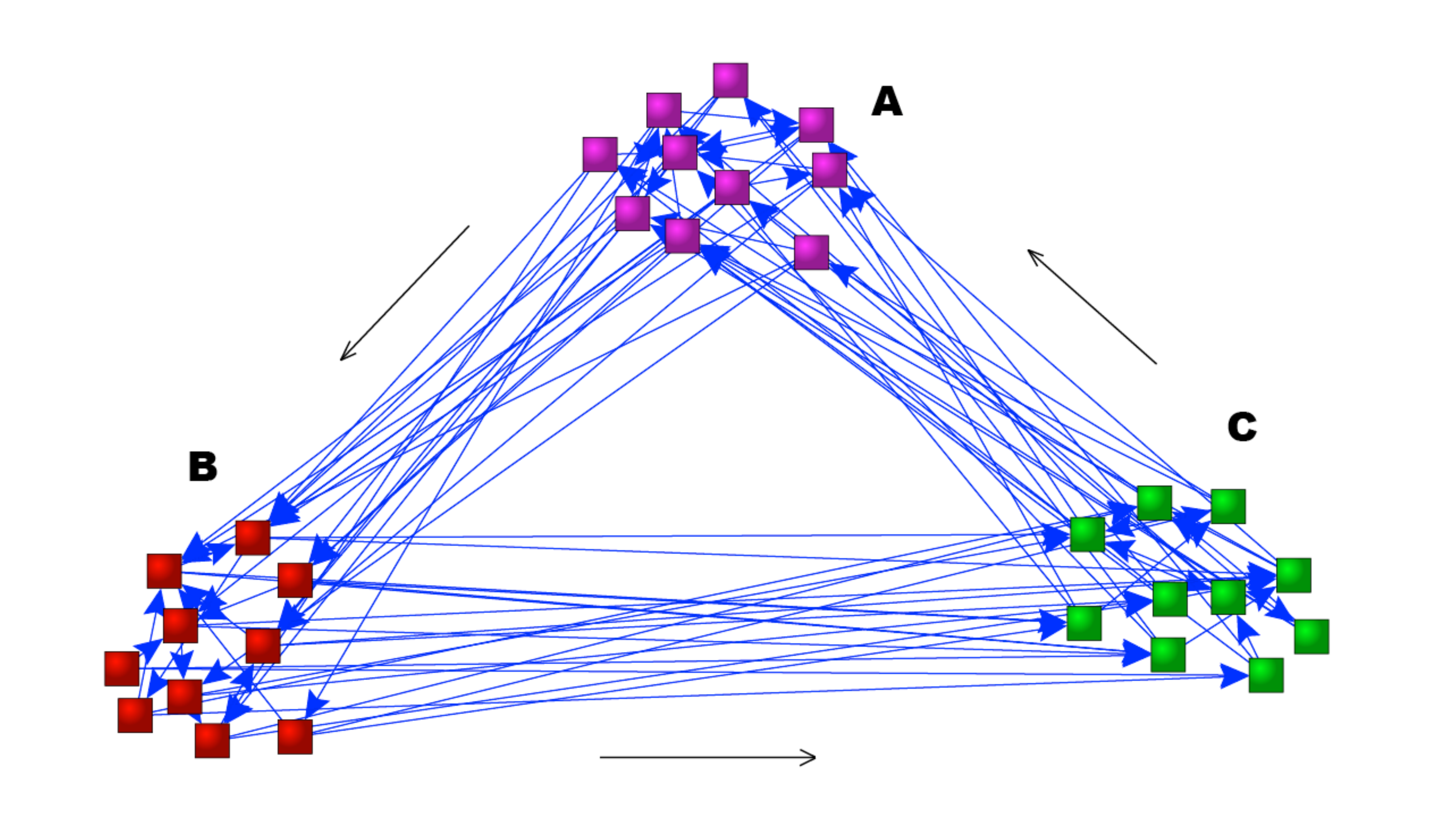} 
\caption {\label{direx} Example of directed graph with a flow running in a cycle between three groups of nodes. 
The directedness of the links enables to distinguish the three groups, and there are methods able to detect them.
Standard community detection methods, instead, are likely to fail. For instance, the value of the directed modularity 
for the partitions in the three groups is zero, whereas the maximum modularity for the graph is positive and corresponds to a different partition.} 
\end{figure}

Methods based on mixture models~\cite{newman07,ramasco08} may detect this kind of structures. Methods 
based on a balance between internal and external links, like (directed) modularity 
optimization may have problems. For example (Fig.~\ref{direx}), consider a network 
with three communities $A$, $B$, $C$, with $10$ nodes in each community, each node with $3$ in-links and $3$ 
out-links on average; nodes in $A$ point to $2$ nodes in $B$, nodes in $B$ point to $2$ nodes in $C$, and 
nodes in $C$ point to $2$ nodes in $A$; each node points to $1$ node in its own community. 
The modularity of this partition is $Q=0$, therefore the optimization would give a different partition,
as the maximum modularity for a graph is usually positive.

\section{Tests}
\label{sec2}

Here we present some tests of community detection methods on our benchmark graphs. 
\begin{figure}[ht]
\centering
\includegraphics[width=\columnwidth]{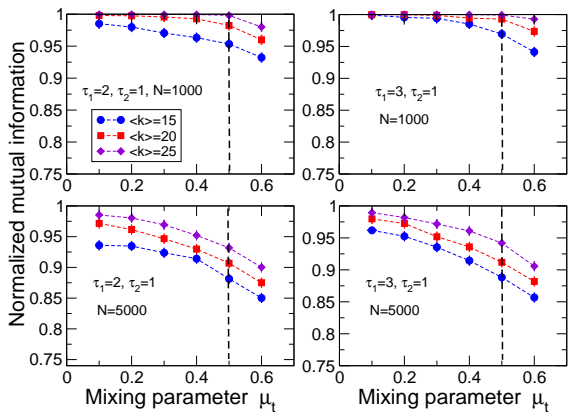} 
\caption{Test of modularity optimization on our benchmark for directed networks. 
The results get worse by increasing the number of nodes and/or decreasing the average degree, 
as we had found for the undirected case in Ref.~\cite{lancichinetti08}. Each point corresponds to an average over $100$ graph realizations.} 
\label{Figdirected}
\end{figure} 
We focused on two techniques: modularity 
optimization, because it is one of very few methods that can be extended to the cases of directed and weighted graphs~\cite{arenas07c};
the Clique Percolation Method (CPM) by Palla et al.~\cite{palla}, a popular method to find community structure with overlapping communities.  
The optimization of modularity was carried out by using simulated annealing~\cite{Guimera:2005}. 
\begin{figure} [ht]
\centering
\includegraphics[height=0.4 \textwidth, width=0.5 \textwidth]{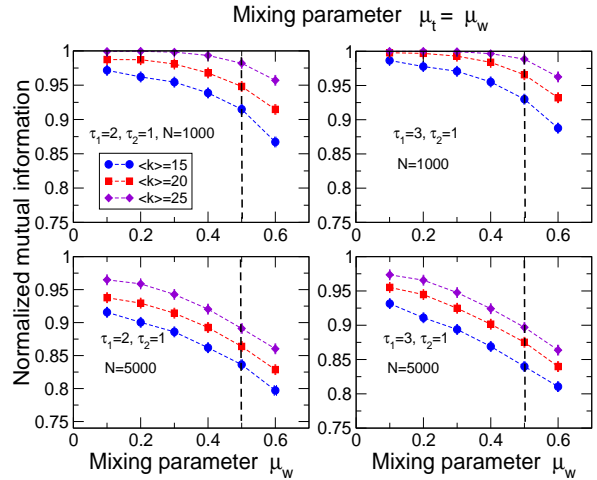} 
\caption{Test of modularity optimization on our benchmark for weighted undirected networks without overlaps between communities. The topological 
mixing parameter $\mu_t$ equals the strength mixing parameter $\mu_w$. 
Each point corresponds to an average over $100$ graph realizations.} 
\label{Figwei1}
\end{figure} 

To measure the similarity between the built-in modular structure of the benchmark 
and the one delivered by the algorithm
we adopt the {\it normalized mutual information}, a measure 
borrowed from information theory~\cite{Danon:2005}. We stress that
other choices for the similarity measure are possible (for a survey, see~\cite{meila}) and that we use the normalized mutual
information for two main reasons: 1) it is regularly used in papers about community detection, so one has a clear
idea of the performance of the algorithms by looking at the results, compared to similar plots; 2) it has been recently extended
to the case of overlapping communities~\cite{lancichinetti09}, whereas most other measures have no such extension.

Fig.~\ref{Figdirected} shows the result for the directed (unweighted) benchmark graphs, without overlapping communities. 
The plot shows a very similar pattern as that observed in the undirected case~\cite{lancichinetti08}.

\begin{figure} [ht]
\centering
\includegraphics[width=\columnwidth]{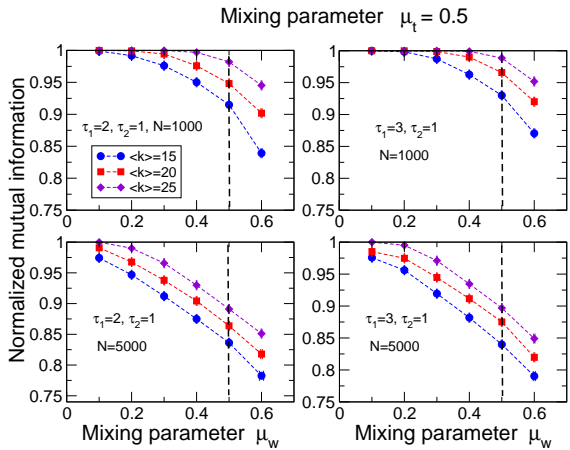} 
\caption {Test of modularity optimization on our benchmark for weighted undirected networks. The topological 
mixing parameter $\mu_t= 0.5$. All other parameters are the same as in Fig.~\ref{Figwei1}.
Each point corresponds to an average over $100$ graph realizations.} 
\label{Figwei2}
\end{figure}

For the weighted benchmark (still without overlapping communities) we can tune two parameters, $\mu_t$ and $\mu_w$. Fig.~\ref{Figwei1} 
refers to networks where we set $\mu_t= \mu_w$, while in Fig.~\ref{Figwei2} we set $\mu_t=0.5$. 
Since, for $\mu_w<0.5$, $\mu_t$ is smaller for the networks of Fig.~\ref{Figwei1} than 
for those in Fig.~\ref{Figwei2}, we would expect to see better performances of modularity optimization in Fig.~\ref{Figwei1} in the range
$0\leq \mu_w< 0.5$. Instead, we get the opposite result. The reason is that the links between communities carry 
on average more weight when $\mu_t<\mu_w$ than when $\mu_t= \mu_w$, and this enhances the chance that mergers between
small communities occur, leading to higher values of modularity~\cite{FB}. Because of such mergers, the partition found by the method
can be quite different from the planted partition of the benchmark.
\begin{figure} [ht]
\centering
\includegraphics[width=\columnwidth]{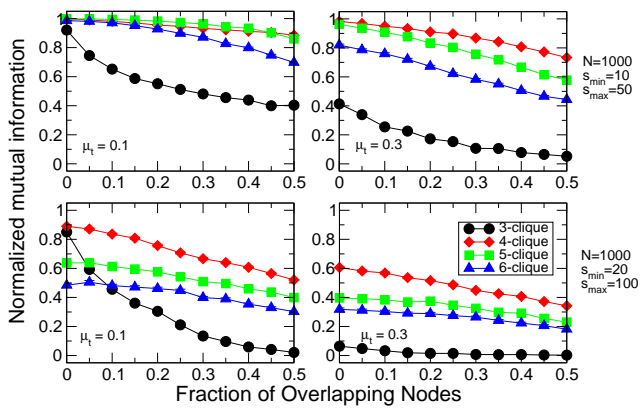} 
\caption {Test of the Clique Percolation Method (CPM) by Palla et al.~\cite{palla} on our benchmark for
undirected and unweighted networks with overlapping communities. The plot shows the variation of the 
normalized mutual information between the planted and the recovered partition, in its generalized form for overlapping communities
~\cite{lancichinetti09}, with the fraction of 
overlapping nodes. The networks have $1000$ nodes, the other parameters are $\tau_1=2$, $\tau_2=1$, 
$\langle k\rangle=20$ and $k_{max}=50$.} 
\label{Figov1}
\end{figure} 
\begin{figure} [ht]
\centering
\includegraphics[width=\columnwidth]{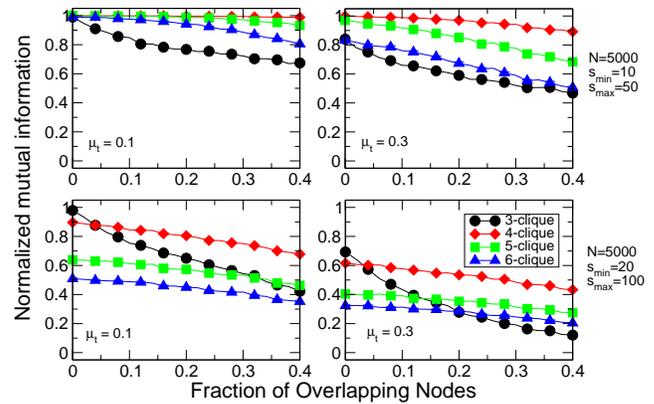} 
\caption{Test of the Clique Percolation Method (CPM) by Palla et al.~\cite{palla} on our benchmark for
undirected and unweighted networks with overlapping communities. The networks have $5000$ nodes, the other parameters are the same 
used for the graphs of Fig.~\ref{Figov1}.} 
\label{Figov2}
\end{figure} 

In Figs.~\ref{Figov1} and \ref{Figov2} we show the results of tests performed with the CPM
on our benchmarks with overlapping communities. In this case, the mixing parameter $\mu_t$ is fixed
and one varies the fraction of overlapping nodes between communities. We have run the CPM for different types of $k$-cliques ($k$ indicates
the number of nodes of the clique), with $k=3,4,5,6$. In general we notice that triangles ($k=3$) yield the worst performance,
whereas $4$- and $5$-cliques give better results. In the two top diagrams community sizes range between $s_{min}=10$ and $s_{max}=50$,
whereas in the bottom diagrams the range goes from $s_{min}=20$ and $s_{max}=100$. By comparing the diagrams in the
top with those in the bottom we see that the algorithm performs better
when communities are (on average) smaller. The networks
used to produce Fig.~\ref{Figov1} consist of $1000$ nodes, whereas those of Fig.~\ref{Figov2} consist of $5000$ nodes.
From the comparison of Fig.~\ref{Figov1} with Fig.~\ref{Figov2} we see that the algorithm performs better on networks of larger size.

\section{Summary}
\label{sec3}

In this paper we have introduced new benchmark graphs to test community detection methods
on directed and weighted networks. The new graphs are suitable extensions of the benchmark we have
recently introduced in Ref.~\cite{lancichinetti08}, in that they account for the fat-tailed distributions
of node degree and community size that are observed in real networks. Furthermore we have equipped all 
our new benchmark graphs with the option of having overlapping communities, an important feature   
of community structure in real networks. With this work we have provided researchers working
on the problem of detecting communities in graphs with a complete set of tools 
to make stringent objective tests of their algorithms, something which is sorely needed in this field.
We have developed and carefully tested a  
software package for the generation of each class of 
benchmark graphs, all of which can be freely downloaded~\cite{software}.

\begin{acknowledgments}

We thank F. Radicchi and J.~J. Ramasco for useful suggestions.

\end{acknowledgments}


\begin{thebibliography}{20}
\bibitem{power} We stress that, since the exponent $\tau_1$ can be arbitrarily chosen, in the limit of large $\tau_1$ one 
converges to a $\delta$-function, so all nodes have the same degree. The same holds for the distribution of community sizes,
which is characterized as well by a tunable power-law exponent $\tau_2$. Therefore, our benchmark is a true generalization
of the benchmark by Girvan and Newman, which is recovered in the limit where the exponents are infinitely large.
\end{thebibliography}
\end{document}